\documentclass{emulateapj}
\usepackage{apjfonts}
\usepackage{color}

\newcommand{\pivec}{\mbox{\boldmath $\pi$}}

\newcommand{\thetae}{\theta_{\rm E}}
\newcommand{\pie}{\pi_{\rm E}}

\def\e{{\rm E}}

\definecolor{darkbrown}{RGB}{139,69,19}

\shorttitle{OGLE-2016-BLG-0263LB}
\shortauthors{HAN, ET AL.}
\begin{document}

\title{
OGLE-2016-BLG-0263L\lowercase{b}: 
Microlensing Detection of a Very Low-mass Binary Companion 
Through a Repeating Event Channel}

\author{
C.~Han$^{01}$, A.~Udalski$^{02,23}$, A.~Gould$^{03,04,05,24}$, 
I.A.~Bond$^{06,25}$\\
and\\
M.~D.~Albrow$^{07}$, S.-J.~Chung$^{03,10}$, Y.~K.~Jung$^{08}$, Y.-H.~Ryu$^{03}$, I.-G.~Shin$^{08}$,
J.~C.,~Yee$^{08}$, W.~Zhu$^{04}$, S.-M.~Cha$^{03,09}$, S.-L.~Kim$^{03,10}$, D.-J.~Kim$^{03}$,
C.-U.~Lee$^{03,10}$, Y.~Lee$^{03,09}$, B.-G.~Park$^{03,10}$   \\
(The KMTNet Collaboration),\\
J.~Skowron$^{02}$, P.~Mr{\'o}z$^{02}$, P.~Pietrukowicz$^{02}$, S.~Koz{\l}owski$^{02}$,
R.~Poleski$^{02,04}$, M.~K.~Szyma{\'n}ski$^{02}$, I.~Soszy{\'n}ski$^{02}$, K.~Ulaczyk$^{02}$,
M.~Pawlak$^{02}$\\
(The OGLE Collaboration) \\
F.~Abe$^{11}$,  Y.~Asakura$^{11}$, R.~Barry$^{12}$, D.P.~Bennett$^{12,13}$, A.~Bhattacharya$^{12,13}$, 
M.~Donachie$^{14}$, P.~Evans$^{14}$, A.~Fukui$^{15}$, Y.~Hirao$^{16}$, Y.~Itow$^{11}$, 
N.~Koshimoto$^{16}$, M.C.A.~Li$^{14}$, C.H.~Ling$^{03}$, K.~Masuda$^{11}$, Y.~Matsubara$^{11}$, 
Y.~Muraki$^{11}$, M.~Nagakane$^{16}$, K.~Ohnishi$^{17}$, C.~Ranc$^{12}$, N.J.~Rattenbury$^{14}$, 
To.~Saito$^{18}$, A.~Sharan$^{14}$, D.J.~Sullivan$^{19}$, T.~Sumi$^{16}$, D.~Suzuki$^{12,20}$, 
P.J.~Tristram$^{21}$, T.~Yamada$^{22}$, T.~Yamada$^{16}$, and A. Yonehara$^{22}$\\
(The MOA Collaboration) 
}

\affil{$^{01}$  Department of Physics, Chungbuk National University, Cheongju 28644, Republic of Korea}
\affil{$^{02}$ Warsaw University Observatory, Al. Ujazdowskie 4, 00-478 Warszawa, Poland}
\affil{$^{03}$ Korea Astronomy and Space Science Institute, Daejon 34055, Republic of Korea}
\affil{$^{04}$ Department of Astronomy, Ohio State University, 140 W. 18th Ave., Columbus, OH 43210, USA}
\affil{$^{05}$ Max Planck Institute for Astronomy, K\"onigstuhl 17, D-69117 Heidelberg, Germany}
\affil{$^{06}$ Institute of Natural and Mathematical Sciences, Massey University, Auckland 0745, New Zealand}
\affil{$^{07}$ University of Canterbury, Department of Physics and Astronomy, Private Bag 4800,
               Christchurch 8020, New Zealand}
\affil{$^{08}$ Smithsonian Astrophysical Observatory, 60 Garden St., Cambridge, MA, 02138, USA}
\affil{$^{09}$ School of Space Research, Kyung Hee University, Yongin 17104, Republic of Korea}
\affil{$^{10}$ Korea University of Science and Technology, 217 Gajeong-ro, Yuseong-gu, Daejeon 34113, 
               Republic of Korea}
\affil{$^{11}$ Institute for Space-Earth Environmental Research, Nagoya University, Nagoya 464-8601, Japan}
\affil{$^{12}$ Code 667, NASA Goddard Space Flight Center, Greenbelt, MD 20771, USA}
\affil{$^{13}$ Department of Physics, University of Notre Dame, Notre Dame, IN 46556, USA}
\affil{$^{14}$ Department of Physics, University of Auckland, Private Bag 92019, Auckland, New Zealand}
\affil{$^{15}$ Okayama Astrophysical Observatory, National Astronomical Observatory of Japan, 3037-5 Honjo, 
               Kamogata, Asakuchi, Okayama 719-0232, Japan}
\affil{$^{16}$ Department of Earth and Space Science, Graduate School of Science, Osaka University, 
               Toyonaka, Osaka 560-0043, Japan}
\affil{$^{17}$ Nagano National College of Technology, Nagano 381-8550, Japan}
\affil{$^{18}$ Tokyo Metropolitan College of Aeronautics, Tokyo 116-8523, Japan}
\affil{$^{19}$ School of Chemical and Physical Sciences, Victoria University, Wellington, New Zealand}
\affil{$^{20}$ Institute of Space and Astronautical Science, Japan Aerospace Exploration Agency, 
               Kanagawa 252-5210, Japan}
\affil{$^{21}$ University of Canterbury Mt.\ John Observatory, P.O. Box 56, Lake Tekapo 8770, New Zealand}
\affil{$^{22}$ Department of Physics, Faculty of Science, Kyoto Sangyo University, 603-8555 Kyoto, Japan}

\footnotetext[23]{The OGLE Collaboration.}
\footnotetext[24]{The KMTNet Collaboration.} 
\footnotetext[25]{The MOA Collaboration.}

\begin{abstract}
We report the discovery of a planet-mass companion to the microlens OGLE-2016-BLG-0263L.  
Unlike most low-mass companions that were detected through perturbations to the smooth 
and symmetric light curves produced by the primary, the companion was discovered through 
the channel of a repeating event, in which the companion itself produced its own single-mass 
light curve after the event produced by the primary had ended.  Thanks to the continuous 
coverage of the second peak by high-cadence surveys, the possibility of the repeating nature 
due to source binarity is excluded with a $96\%$ confidence level.  The mass of the companion 
estimated by a Bayesian analysis is $M_{\rm p}=4.1_{-2.5}^{+6.5}\ M_{\rm J}$.  The projected 
primary-companion separation is $a_\perp = 6.5^{+1.3}_{-1.9}$ au.  The ratio of the separation 
to the snow-line distance of $a_\perp/a_{\rm sl}\sim 15.4$  corresponds to the region beyond 
Neptune, the outermost planet of the solar system.  We discuss the importance of high-cadence 
surveys in expanding the range of microlensing detections of low-mass companions and future 
space-based microlensing surveys.
\end{abstract}

\keywords{gravitational lensing: micro -- planetary systems -- brown dwarfs}

\section{Introduction}
 
A microlensing signal 
of a very low-mass companion such as a planet is usually a brief 
perturbation to the smooth and symmetric lensing light curve produced by the single 
mass of the primary lens.  Short durations of perturbations combined with the non-repeating 
nature of lensing events imply that microlensing detections of low-mass companions require 
high-cadence observations. During the first decade of microlensing surveys when the survey 
cadence was not sufficiently high to detect short companion signals, lensing experiments 
achieved the required observational cadence by employing a strategy 
in which lensing events were detected by wide-field surveys and 
a fraction of these events 
were monitored using 
multiple narrow-field telescopes \citep{Gould1992b, Udalski2005, Beaulieu2006}.

Thanks to the instrumental upgrade of existing surveys and the addition of new surveys, 
the past decade has witnessed a great increase of the observational cadence of lensing 
surveys. By entering the fourth phase survey experiment, the Optical Gravitational 
Lensing Experiment (OGLE) group substantially increased the observational cadence by 
broadening the field of view (FOV) of their camera from 0.4 ${\rm deg}^2$ to 1.4 
${\rm deg}^2$ \citep{Udalski2015}. In addition, the Korea Microlensing Telescope 
Network (KMTNet) group started a microlensing survey in 2015 using 3 globally distributed 
telescopes each of which is equipped with a camera having 4 ${\rm deg}^2$ FOV \citep{Kim2016}. 
Furthermore, the Microlensing Observation in Astrophysics (MOA) group \citep{Bond2001, Sumi2003} 
plans to add a new infrared telescope (T. Sumi 2017, private 
communication) into the survey. With the elevated sampling rate, microlensing surveys have 
become increasingly capable of detecting short signals without the need of followup observations, 
e.g.\ OGLE-2012-BLG-0406Lb \citep{Poleski2014b}, OGLE-2015-BLG-0051/KMT-2015-BLG-0048Lb 
\citep{Han2016}, OGLE-2016-BLG-0954Lb \citep{Shin2016}, and OGLE-2016-BLG-0596Lb \citep{Mroz2017}.

One most important merit of high-cadence microlensing surveys is the increased rate of detecting 
very low-mass companions. Currently, more than 2000 lensing events are being detected every 
season.  Due to the limited resources, however, only a handful events can be monitored by 
followup observations. In principle, followup observations can be started at the early stage 
of anomalies, but implementing this strategy in practice is challenging due to the difficulty 
in detecting short anomalies in their early stages.  On the other hand, high-cadence surveys 
are capable of continuously and densely sampling light curves of all microlensing events, and 
thus the rate of detecting very low-mass companions is expected to be greatly increased.

Another important advantage of high-cadence surveys is that they open an additional channel 
of detecting very low-mass companions.  By definition, under the survey+followup strategy, 
events can only be densely monitored by followup observations once they have been alerted by 
surveys. Furthermore, followup resources are limited, so in practice those observations have 
been confined to those located in the narrow region of separations from the host star, the 
so-called `lensing zone' \citep{Gould1992b, Griest1998}.  On the other hand, high-cadence 
surveys enable to densely monitor events not only during the lensing magnification but also 
before and after the lensing magnification, and this allows low-mass companions to be detected 
via the `repeating-event' channel.  The signal through the repeating-event channel is produced 
by a companion with a projected separation substantially larger than the Einstein radius of the 
primary star and it occurs when the source trajectory passes the effective magnification regions 
of both the primary star and the companion \citep{Distefano1999}.  Thus, the two lenses (primary 
and companion) act essentially independently and appear to give rise to two separate microlensing 
events with different time scales (related by the square root of their mass ratio) but the same 
source star.  Therefore, the channel is important because it expands the region of microlensing 
detections of low-mass companions to larger separations.  Under the assumption of power-law 
distributions of host-planet separations, \citet{Han2007} estimated that planets detectable 
by high-cadence surveys through the repeating channel will comprise $\sim 3$ -- 4\% of all 
planets.

In this paper, we report the discovery of a planet-mass binary companion through the 
repeating-event channel.  In Section 2 , we describe the survey observations that led 
to the discovery of the companion.  In Section 3, we explain the procedure of analyzing 
the observed lensing light curve and present the physical parameters of the lens system.  
We discuss the importance of the repeating-event channel in Section~4.

\begin{figure}
\includegraphics[width=\columnwidth]{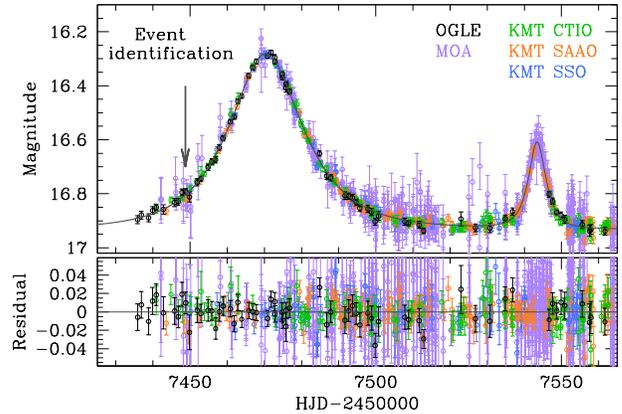}
\caption{Light curve of OGLE-2016-BLG-0263.  The curve superposed on the data points 
represents the best-fit binary-lens model.  The arrow denotes the time when the event 
was first discovered.  The lower panel shows the residual from the model.
}
\label{fig:one}
\end{figure}

\section{Observation and Data}

The low-mass binary companion was discovered from the observation of the microlensing event 
OGLE-2016-BLG-0263.  In Figure~\ref{fig:one}, we present the light curve of the event. The 
event occurred on a star located toward the Galactic bulge field with equatorial coordinates 
$({\rm RA},{\rm DEC})_{\rm J2000}= (17^\circ 59'34''\hskip-2pt.9, -31^{\rm h}49^{\rm m}07^{\rm s}\hskip-2pt.0)$
 that are equivalent to the Galactic coordinates $(l,b)=(-0^\circ\hskip-2pt .95, -4^\circ\hskip-2pt.06)$. 
The lensing-induced brightening of the source star was identified on 2016 March 1 
(${\rm HJD}'={\rm HJD}-2450000=7448.7$) by the Early Warning System of the OGLE survey 
\citep{Udalski1994, Udalski2003} using the 1.3m Warsaw telescope at the Las Campanas Observatory 
in Chile. Observations by the OGLE survey were conducted with a $\sim 1$ day cadence, and most 
images were taken in the standard Cousins $I$ band with occasional observations in the Johnson 
$V$ band for color measurement. After being identified, the event followed a standard point-source 
point-lens (PSPL) light curve, peaked at ${\rm HJD}'\sim 7470$, and gradually returned to the 
baseline magnitude of $I\sim 16.9$.

However, after returning to baseline, the source began to brighten again. The anomaly was noticed on 
2016 May 30 (${\rm HJD}\sim 7538$) and announced to the microlensing community for possible followup 
observations although no followup observation was conducted. The anomaly, which continued about 
10 days, appears to be an independent PSPL event with a short time scale. The time between 
the first and second peaks of the light curve is $\sim 73$ days.

The event was also in the footprint of the KMTNet and MOA
surveys. The survey utilizes three globally 
distributed 1.6m telescopes that are located at the Cerro Tololo Interamerican Observatory in 
Chile (KMTC), the South African Astronomical Observatory in South Africa (KMTS), and the Siding 
Spring Observatory in Australia (KMTA).  Similar to OGLE observations, most of the KMTNet data 
were acquired using the standard Cousins $I$-band filter with occasional $V$-band observations. 
The event was in the BLG34 field for which observations were carried out with a $\sim 2.5$ hr 
cadence. 
The MOA survey uses the 1.6 m telescope located at the Mt.~John University Observatory 
in New Zealand. Data were acquired in a customized $R$ band filter with a bandwidth 
corresponding to the sum of the Cousin $R$ and $I$ bands. 
The event was independently found by the MOA survey and was dubbed MOA-2016-BLG-075.

\begin{deluxetable}{lrr}
\tablecaption{Error bar correction factors \label{table:one}}
\tablewidth{0pt}
\tablehead{
\multicolumn{1}{c}{Data set} &
\multicolumn{1}{c}{$k$}  &
\multicolumn{1}{c}{$\sigma_{\rm min}$}  
}
\startdata                                              
OGLE            &  1.452    &   0.001     \\     
MOA             &  1.212    &   0.001     \\     
KMT (CTIO)      &  1.204    &   0.001     \\ 
KMT (SAAO)      &  1.806    &   0.001     \\      
KMT (SSO)       &  1.300    &   0.001       
\enddata                                              
\end{deluxetable}

Photometry of the images was conducted using pipelines based on the Difference 
Imaging Analysis method \citep{Alard1998, Wozniak2000} and customized by the individual groups: 
\citet{Udalski2003} for the OGLE, \citet{Albrow2009} for the KMTNet, and 
\citep{Bond2001} for the MOA groups. 
In order to 
analyze the data sets acquired by different instruments and reduced by different photometry 
pipelines, we readjust error bars of the individual data sets. Following the usual procedure 
described in \citet{Yee2012}, we normalize the error bars by
\begin{equation}
\sigma=k (\sigma_0^2+\sigma_{\rm min}^2)^{1/2},
\label{eq1}
\end{equation}
where $\sigma_0$ is the error bar estimated from the photometry pipeline, $\sigma_{\rm min}$ is
a term used to adjust error bars to be consistent with the scatter of the data set, and $k$ is 
a normalization factor used to make the $\chi^2$ per degree of freedom unity. The $\chi^2$ 
value is computed based on the best-fit solution of the lensing parameters obtained from modeling 
(Section 3). In Table~\ref{table:one}, we list the error-bar adjustment factors for the individual 
data sets.  We note that the OGLE data used in our analysis were rereduced for optimal photometry 
and error bars were estimated according to the prescription described in \citet{Skowron2016},
although one still needs a non-unity ($k\neq 1$) scaling factor to make $\chi^2/{\rm dof}=1$.

\begin{figure}
\includegraphics[width=\columnwidth]{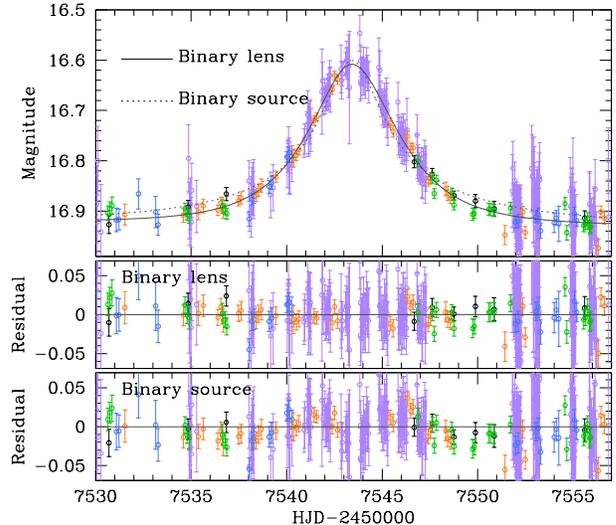}
\caption{Enlarged view of the light curve around the second peak. Superposed on the 
data points are the model light curves obtained from binary-lens (solid) and binary-source 
(dotted) analysis. The lower panels show the residual from the individual models.
}
\label{fig:two}
\end{figure}

\section{Analysis}

The light curve of OGLE-2016-BLG-0263 is characterized by two peaks in which the short 
second one occurred well after the first one. 
The light curve of such a repeating event can be produced in two cases. The first case 
is a binary-source event in which  the double peaks are produced when the lens passes 
close to both components of the source separately, one after another 
\citep{Griest1992, Sazhin1994, Han1997}.  The other case is a binary-lens event 
where the source approaches both components of a widely separated 
binary lens, and the source flux is successively magnified by the individual lens components 
\citep{Distefano1996}.  The degeneracy between binary-source and binary-lens perturbations was 
first discussed by \citet{Gaudi1998}.  In order to investigate the nature of the second peak, 
we test both the binary-source and binary-lens interpretations.

\subsection{Binary-Source Interpretation}

The light curve of a repeating binary-source event is represented by the superposition of 
the PSPL light curves involved with the individual source stars, i.e.
\begin{equation}
A_{\rm BS}={A_1F_{0,1}+A_2F_{0,2}   \over F_{0,1}+F_{0,2} }
={A_1+A_2q_F \over 1+q_F }.
\label{eq2}
\end{equation}
Here $F_{0,i}$ represents the baseline fluxes of the individual source components and 
$q_F=F_{0,2}/F_{0,1}$ is the flux ratio between the source components. The lensing magnification 
involved with each source component is represented by
\begin{equation}
A_i={u_i^2+2  \over u_i(u_i^2+4)^{1/2}};\qquad 
u_i = \left[ u_{0,i}^2+ \left( { t-t_{0,i}\over t_{\rm E}} \right)^2 \right]^{1/2},
\label{eq3}
\end{equation}
where $t_{0,i}$ is the time of the closest lens-source approach, $u_{0,i}$ is the lens-source 
separation at that moment, and $t_{\rm E}$ is the Einstein time scale. For the basic description 
of the light curve of a binary-source event, therefore, one needs 6 lensing parameters including 
$t_{0,1}$, $t_{0,2}$, $u_{0,1}$, $u_{0,2}$, $t_{\rm E}$, and $q_F$ \citep{Hwang2013}.
The light curve is then modeled as
\begin{equation}
F_j(t_k) = F_{s,j}A_{\rm BS}(t_k;t_{0,1},u_{0,1},t_{0,2},u_{0,2},t_{\rm E},q_F), + F_{b,j},
\label{eq4}
\end{equation}
where the $(F_{s,j},F_{b,j})$ are specified separately for each observatory
but there is a single $q_F$ for all observatories using a single band
(e.g., $I$ band).

We model the observed light curve based on the binary-source parameters. Since the light 
curve of a binary-source event varies smoothly with the changes of the lensing parameters, 
we search for the best-fit parameters by $\chi^2$ minimization using a downhill approach. 
For the downhill approach, we use the Markov Chain Monte Carlo (MCMC) method. We set the 
initial values of $t_{0,1}$ and $t_{0,2}$ based on the times of the first and second peaks, 
respectively, while the initial values of $u_{0,1}$ and $u_{0,2}$ are determined based on 
the peak magnifications of the individual peaks. Since both PSPL curves of the individual 
peaks share a common time scale\footnote{In the Appendix, we discuss the possibility of 
different time scales due to the orbital motion of the source.}, we set the initial value 
of $t_{\rm E}$ as the one estimated based on the PSPL fitting of the light curve with the 
first peak.  The initial value of the flux ratio $q_F$ is guessed based on the values of 
$u_{0,i}$.

\begin{deluxetable}{lr}
\tablecaption{Best-fit binary-source solution \label{table:two}}
\tablewidth{0pt}
\tablehead{
\multicolumn{1}{c}{Parameter}  &
\multicolumn{1}{c}{Value}  
}
\startdata                                              
$\chi^2$             &    2598.8                 \\     
$t_{0,1}$ (HJD)      &  2457470.441 $\pm$ 0.028  \\ 
$t_{0,2}$ (HJD)      &  2457543.426 $\pm$ 0.028  \\      
$u_{0,1}$            &  0.646 $\pm$ 0.032        \\    
$u_{0,2}$            &  0.095 $\pm$ 0.004        \\ 
$t_{\rm E}$ (days)   &  15.33 $\pm$ 0.50         \\      
$q_{F,I}$            &  0.037 $\pm$ 0.002        \\
$q_{F,R}$            &  0.036 $\pm$ 0.002        \\
$F_s/F_b$            &  2.452/0.219
\enddata                                              
\end{deluxetable}

In Table~\ref{table:two}, we present the parameters of the best-fit binary-source solution. 
Also presented is the ratio of the source flux $F_s$ to that of the blend $F_b$ that are 
estimated from the OGLE data set.  The uncertainties of the lensing parameters are estimated 
based on the scatter of points on the MCMC chain.  According to the solution, the second peak 
was produced by the lens approaching very close to the second source which is approximately 
30 times fainter than the primary source star.  In Figure~\ref{fig:two}, we also present the 
model light curve (dotted curve) superposed on the observed data points. At first glance, the 
model appears to describe the overall shape of the second peak. However, careful inspection 
of the model light curve and the residual reveals that the fit is inadequate not only in the 
rising and falling parts but also near the peak part of the light curve.

We check whether the fit can be further improved with higher-order effects. 
The trajectory of the lens with respect to the source 
might deviate from rectilinear due to the orbital motion of the Earth around the sun. 
We check 
this so-called `microlens-parallax' effect \citep{Gould1992a} by conducting additional modeling. 
Accounting for microlens-parallax effects requires to include 2 additional parameters of 
$\pi_{{\rm E},N}$ and $\pi_{{\rm E},E}$, which represent the components of the microlens 
parallax vector $\pivec_{\rm E}$ projected onto the sky along the north and east equatorial 
coordinates, respectively. The direction of $\pivec_{\rm E}$ corresponds to that of the 
relative lens-source motion in the Earth's frame. The magnitude of 
$\pivec_{\rm E}$ is $\pi_{\rm E}=\pi_{\rm rel}/\theta_{\rm E}$, where 
$\pi_{\rm rel}={\rm au}(D_{\rm S}^{-1}-D_{\rm L}^{-1})$ is the relative lens-source parallax and
$D_{\rm _L}$ and $D_{\rm S}$ represent the distances to the lens and source, respectively. 
From the modeling with parallax effects, we find that the improvement of the fit is very 
minor with $\Delta\chi^2\sim 4.4$.

\begin{figure}
\includegraphics[width=\columnwidth]{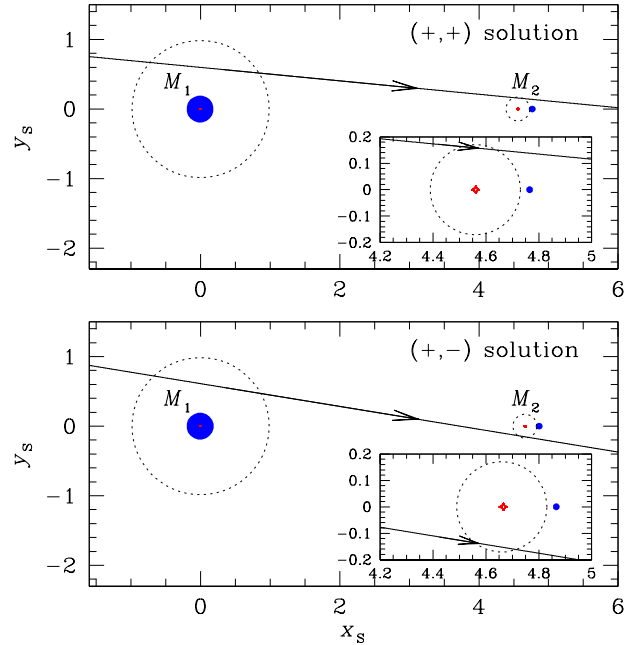}
\caption{Lens system geometry that shows the source trajectory (line with an arrow) with 
respect to the binary-lens components (blue dots). $M_1$ and $M_2$ denote the heavier and 
lower-mass components of the binary lens. The dotted circles represent the boundary of 
effective lensing magnification and the size of each circle corresponds to the Einstein 
radius corresponding to the mass of each lens component. The tiny close curves at the centers 
of the dotted circles represent the caustics. The inset shows the enlarged view of the 
caustic located close to $M_2$.
}
\label{fig:three}
\end{figure}

\subsection{Binary-Lens Interpretation}

Unlike the case of a binary-source event, the light curve of a binary-lens event cannot be 
described by the superposition of the two light curves involved with the individual lens 
components because the lens binarity induces a region of discontinuous 
lensing magnifications, i.e. caustics. As a result, the lensing parameters needed to describe 
a binary-lens event is different from those of a binary-source event. Basic description of 
a binary-lens event requires 7 principal parameters. The first three of these parameters, 
$t_0$, $u_0$, and $t_{\rm E}$, are the same as those of a single-lens event. The other three 
parameters describe the binary lens including the projected separation $s$ (normalized to 
$\theta_{\rm E}$) and the mass ratio $q$ between the binary components, and the angle between 
the source trajectory and the binary axis, $\alpha$. Light curves produced by binary lenses 
are often identified by characteristic spike features that are produced by the source 
crossings over or approaches close to caustics. In this case, the caustic-involved parts of 
the light curve are affected by finite-source effects. To account for finite-source effects, 
one needs an additional parameter $\rho=\theta_*/\theta_{\rm E}$, where $\theta_*$ is the 
angular source radius. For OGLE-2016-BLG-0263, however, the light curve does not show any 
feature involved with a caustic and thus we do not include $\rho$ as a parameter.

Binary lenses form caustics of 3 topologies \citep{Schneider1986, Erdl1993}, which are usually 
referred to as `close', `resonant', and `wide'. For a `resonant' binary, where the projected 
binary separation is equivalent to the angular Einstein radius, i.e.\ $s\sim 1$, the caustics 
form a single big closed curve with 6 cusps. For a `close' binary with $s< 1-3q^{1/2}/4$ 
\citep{Dominik1999}, the caustic consists of two parts, where one four-cusp caustic is located 
around the barycenter of the binary lens and two small three-cusp caustics are positioned away 
from the barycenter. For a `wide' topology with $s > 1+3q^{1/2}/2$ \citep{Dominik1999},  
there exist two four-cusp caustics which are located close to the individual lens components.

A repeating binary-lens event is produced by a wide binary lens, and the individual peaks of the 
repeating event occur when the source approaches the four-cusp caustics of the wide binary lens. 
The caustic has an offset of $\Delta x \sim q/{s(1+q)}$ with respect to each lens position toward 
the other lens component \citep{Distefano1996, An2002}.  In the very wide binary regime with 
$s \gg 1$, each of the two caustics is approximated by the tiny astroidal Chang-Refsdal caustic 
with an external shear $\gamma=q/[s^2(1+q)]$ \citep{Chang1984} and the offset $\Delta x \rightarrow 0$, 
implying that the position of the caustic approaches that of the lens components.  In this regime, 
the light curves involved with the individual binary-lens components are described by two separate 
PSPL curves, and the light curve of the repeating event is approximated by the superposition of the 
two PSPL curves,  i.e.\ $F_{\rm obs}(t)=F_S[A_1(t)+A_2(t)]+F_b$, where $F_{\rm obs}$ is the observed 
flux and $A_1$ and $A_2$ represent the lensing magnifications involved with the individual lens 
components.  To be noted is that the time scales of the two PSPL curves of a repeating event are 
proportional to the square root of the masses of the lens components, i.e.\ 
$t_{{\rm E},2}/t_{{\rm E},1}=(m_2/m_1)^{1/2}=q^{1/2}$, while the time scales of the two PSPL curves 
of a repeating binary-source event are the same because both PSPL curves are produced by a common lens.

To test the binary-lens interpretation, we conduct binary-lens modeling of the observed 
light curve.  Similar to the binary-source case, we set the initial values of the lensing 
parameters based on the time of the major peak for $t_0$, the peak magnification of the 
major event for $u_0$, the duration of the major event for $t_{\rm E}$, the ratio of the 
time gap between the two peaks to the event time scale for $s\sim \Delta t/t_{\rm E}$, the 
ratio between the time scales of the first and second events for 
$q\sim (t_{{\rm E},2}/t_{{\rm E},1})^2$, and $\alpha\sim 0$ for 
a repeating binary-lens event. Based on these initial values, we search for a binary-lens 
solution using the MCMC downhill approach. To double check the result, we conduct a grid 
search for a solution in the parameter space of $(s, q, \alpha)$. From this, we confirm 
that the solution found based on the initial values of the lensing parameters converges to 
the solution found by the grid search.

Although the binary-lensing model does not suffer from the degeneracy in the $s$ and $q$ 
parameters, it is found that there exists a degeneracy in the source trajectory angle 
$\alpha$.  This degeneracy occurs because a pair of solutions with source trajectories 
passing the lens components on the same, (+,+) solution, and the opposite, (+,-) solution, 
sides with respect to the binary axis result in similar light curves.  See Figure~\ref{fig:three}.
For OGLE-2016-BLG-0263, we find that the (+,+) solution is 
slightly preferred over the (+,-) solution by $\Delta\chi^2=7.8$.

\begin{deluxetable}{lrr}
\tablecaption{Best-fit binary-lens solution \label{table:three}}
\tablewidth{0pt}
\tablehead{
\multicolumn{1}{c}{Parameter}  &
\multicolumn{1}{c}{(+,+) solution}      &
\multicolumn{1}{c}{(+,-) solution}       
}
\startdata                                              
$\chi^2$           &  2438.2                  &   2446.0                         \\   
$t_0$ (HJD)        &  2457470.433 $\pm$ 0.036 &   2457470.432 $\pm$ 0.036        \\ 
$u_0$              &  0.581 $\pm$ 0.027       &       0.599  $\pm$  0.031        \\      
$t_{\rm E}$ (days) &  16.24 $\pm$ 0.45        &      15.92   $\pm$  0.51         \\    
$s$                &  4.72 $\pm$ 0.12         &       4.86  $\pm$   0.15         \\ 
$q$ ($10^{-2}$)    &  3.06 $\pm$ 0.08         &       2.97  $\pm$   0.09         \\      
$\alpha$ (radian)  &  0.095 $\pm$ 0.002       &       0.163  $\pm$  0.003        \\
$F_s/F_b$          &  2.419/0.254             &       2.543/0.131  
\enddata                                              
\end{deluxetable}

In Table~\ref{table:three}, we present the best-fit binary-lens parameters along with 
the $\chi^2$ value of the fit.  Since the degeneracy between (+,+) and (+,-) solutions 
is quite severe, we present both solutions.  Because the difference between the source trajectory 
angles of the two solutions is small, it is found that the lensing parameters of the two 
solutions are similar to each other.  Two factors to be noted are first the binary separation, 
$s\sim 4.7$, is substantially greater than the Einstein radius and second the mass ratio 
between the lens components, $q\sim 0.03$, is quite small.  We 
present the model light curve of the best-fit binary-lens solution, i.e.\ (+,+) solution, 
in Figure~\ref{fig:one} for the whole event and in Figure~\ref{fig:two} for the second peak.

In Figure~\ref{fig:three}, we present the lens system geometry that shows the source 
trajectory (line with an arrow) with respect to the lens components (marked by blue dots).
The upper and lower panels are for the (+,+) and (+,-) solutions, respectively.
The tiny red cuspy closed curves near the individual lens components represent the caustics. 
We note that all lengths are scaled to the angular Einstein radius corresponding to the 
total mass of the binary lens. The two dotted circles around the individual caustics represent 
the Einstein rings corresponding to the masses of the individual binary-lens components with 
radii $r_1=[1/(1+q)]^{1/2}$ and $r_2=[q/(1+q)]^{1/2}$. From the geometry, one finds that 
the source trajectory approached both lens components and the two peaks in the lensing 
light curve were produced at the moments when the source approached the caustics near 
the individual lens components. In the regime with a small mass ratio, $q\ll 1$, the 
caustics located close to the higher and lower-mass lens components are often referred 
to as `central' and `planetary' caustics, respectively. The small central caustic is located 
very close to the higher-mass lens component and its size as measured by the width along the 
binary axis is $\sim 4q/(s-s^{-1})^2\sim 0.006$ \citep{Chung2005}. The comparatively 
larger planetary caustic is located on the side of the lower-mass lens component with a 
separation from the heavier lens component of $\sim s-1/s\sim 4.6$. The size of the 
planetary caustic is related to the separation and mass ratio of the binary lens by 
$\sim 4q^{1/2}/[s(s^2-1)^{1/2}]\sim 0.03$ \citep{Han2006}. Since the distance to each 
caustic from the source trajectory is much greater than the caustic size, 
the light curve involved with each lens component appears as a PSPL curves.

\begin{figure}
\includegraphics[width=\columnwidth]{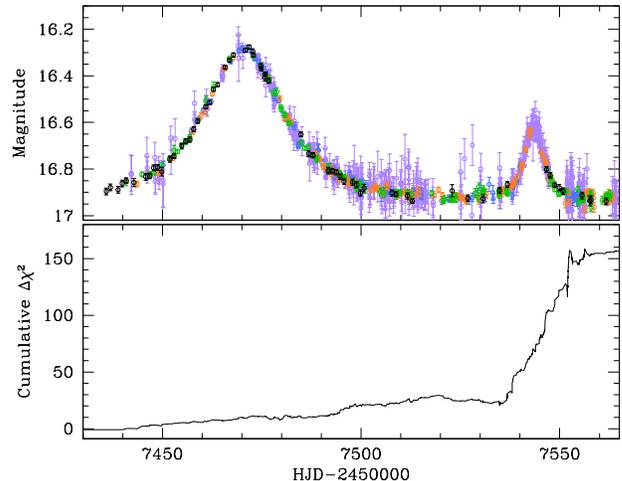}
\caption{Cumulative distribution of $\Delta\chi^2=\chi_{\rm BS}^2-\chi_{\rm BL}^2$, 
where $\chi_{\rm BS}^2$ and $\chi_{\rm BL}^2$ represent the $\chi^2$ values of the 
binary-source and binary-lens models, respectively.
}
\label{fig:four}
\end{figure}

\subsection{Comparison of Models}

Knowing that both binary-source (BS) and binary-lens (BL) interpretations can explain the 
repeating nature of the lensing light curve, we compare the two models in order to find 
the correct interpretation of the event. For this, we construct the cumulative distribution 
of $\chi^2$ difference between the two models.

Figure~\ref{fig:four} shows the constructed $\Delta\chi^2$ distribution where 
$\Delta\chi^2=\chi_{\rm BS}^2-\chi_{\rm BL}^2$. The distribution shows that the 
binary-lens interpretation better describes the observed light curve than the 
binary-source interpretation does. The biggest $\Delta\chi^2$ occurs during the 
second peak. This can be seen also in Figure~\ref{fig:two}, where the residuals 
from both models around the second peaks are presented. The total $\chi^2$ difference 
is $\Delta\chi^2\sim 160$. 
To show the statistical significance of the difference between the two models, we 
conduct a $F$-test for the residuals from the models in the region around the 
second peak.
From this, we find $F=1.78$. 
This corresponds to a $\sim 96\%$ probability that the two models have different variances, 
suggesting that the models can be distinguished with a significant confidence level.

We note that the unambiguous discrimination between the two interpretations was possible 
due to the continuous coverage of the second peak using the globally distributed telescopes. 
One may note large gaps in the observations from Chile ($7537 < {\rm HJD}' < 7546$) and
Australia ($7540 < {\rm HJD}' < 7551$), which were both due to bad weather.  Nevertheless, 
the anomaly was continuously covered by the 
KMTS and MOA data,
enabling accurate interpretation of the event.

\begin{figure}
\includegraphics[width=\columnwidth]{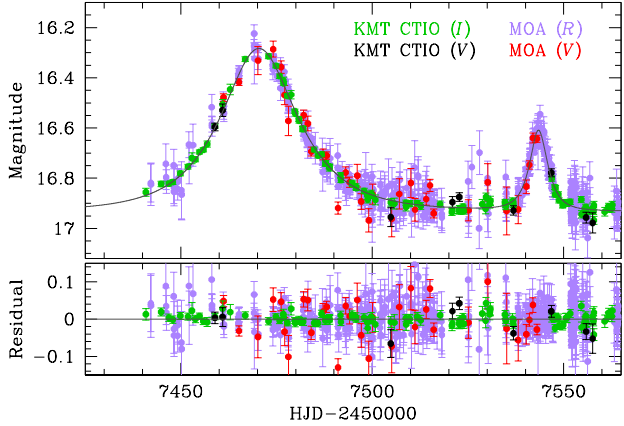}
\caption{
$V$-band data from the MOA and KMTNet surveys.
}
\label{fig:five}
\end{figure}

Another way to discriminate the binary-source/binary-lens interpretations is to use 
color information. This is possible because the color measured during the two peaks 
would be different for the binary-source interpretation while the colors should be 
the same for the binary-lens interpretation. 
According to the small flux ratio presented in Table~\ref{table:two}, the 
stellar types of the source stars would be greatly different.
If a binary-source interpretation is correct, then, the source stars 
should have significantly different colors. 
The second peak was observed in $V$ band by the MOA and KMTNet surveys.  In 
Figure~\ref{fig:five}, we present the $V$-band data plotted over the $I$ and 
$R$-band data,  showing that the second peak was covered in $V$ band 
with 6 and 2 points by the  MOA and KMTNet surveys, respectively.  
In the binary-source modeling, we introduce two flux ratios $q_{F,I}$ and $q_{F,R}$ 
to check the possibility of measuring the color difference between the source stars, i.e.\ 
$\Delta(R-I)=(R-I)_1-(R-I)_2 = 2.5 \log (q_{F,I}/q_{F,R})$. 
We note that the $R$-band flux ratio is measured based on the MOA data. 
From this, we find 
$q_{F,I}=0.037 \pm 0.02$ and $q_{F,R}=0.036 \pm 0.02$, indicating 
no color change within the error bar. 
This suggests the inconsistency in the binary-source interpretation and further 
supports the binary-lens interpretation.

\subsection{Source Star}

Characterizing the source star of a lensing event is important for caustic-crossing binary-lens 
events because the angular source radius $\theta_*$ combined with the normalized source radius 
$\rho$ enables one to determine the angular Einstein radius, i.e. $\theta_{\rm E}=\theta_*/\rho$. 
Although one cannot determine $\theta_{\rm E}$ for OGLE-2016-BLG-0263 because the source did not 
cross caustics and thus the light curve is not affected by finite-source effects, we characterize 
the source star for the sake of completeness.

The source star is characterized based on its de-reddened color $(V-I)_0$ and brightness $I_0$. 
We determine $(V-I)_0$ and $I_0$ of the source star using the usual method of \citet{Yoo2004}, 
where the instrumental color and brightness of the source are calibrated using the position of 
the giant clump (GC) centroid, for which the de-reddened color and brightness 
$(V-I,I)_{0,{\rm GC}}=(1.06,14.63)$ \citep{Bensby2011, Nataf2013} are known.

\begin{figure}
\includegraphics[width=\columnwidth]{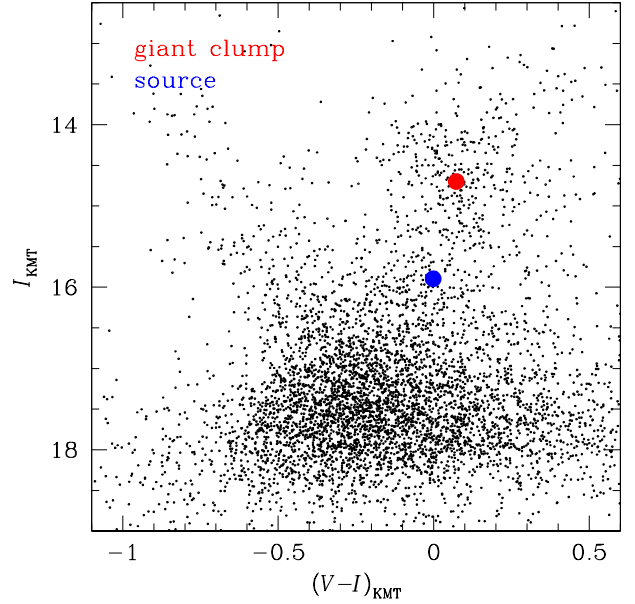}
\caption{Position of the source star with respect to the centroid of giant clump in the 
instrumental color-magnitude diagram of stars in the neighboring region around the source.
}
\label{fig:six}
\end{figure}

Figure~\ref{fig:six} 
shows the position of the source star with respect to the GC centroid in the instrumental 
color-magnitude diagram of stars in the $205"\times 205"$ image stamp centered at the 
source position.  The locations of the source and GC centroid are $(V-I)=(-0.07,15.89)$ 
and $(V-I)_{\rm GC}=(0.07, 14.70)$, respectively.  From the offsets in color 
$\Delta(V-I)=(V-I)-(V-I)_{\rm GC}=-0.14$ and magnitude $\Delta I=I-I_{\rm GC}=1.19$, we 
estimate that the re-reddened color and magnitude of the source star are
 $(V-I,I)_0=(0.99, 15.82)$.
This indicates that the source is a K-type giant star.

\subsection{Physical Parameters}

For the unique determination of the mass $M$ and distance $D_{\rm L}$ to the lens, 
one needs to measure both the microlens parallax $\pi_{\rm E}$ and the angular 
Einstein radius $\theta_{\rm E}$ that are related to $M$ and $D_{\rm L}$ by
\begin{equation}
M={\thetae \over \kappa \pie};\qquad
D_{\rm L}={{\rm au} \over \pie\thetae+\pi_{\rm S}},
\label{eq5}
\end{equation}
where $\kappa\equiv4G/(c^2\ {\rm au})\simeq 8.144\ {\rm mas}\ M_\odot^{-1}$ and 
$\pi_{\rm S}$ denotes the source parallax. 
For OGLE-2016-BLG-0263, none of these quantities is measured and thus the physical 
parameters cannot be uniquely determined. However, one can still statistically constrain 
the physical lens parameters based on the measured event time scale $t_{\rm E}$ that is 
related to the physical parameters by
\begin{equation}
t_{\rm E}={(\kappa M \pi_{\rm rel})^{1/2}\over \mu};\qquad
\pi_{\rm rel}={\rm au}\left({1\over D_{\rm L}} - {1\over D_{\rm S}}\right),
\label{eq6}
\end{equation}
where $\mu$ represents the relative lens-source proper motion.

In order to estimate the mass and distance to the lens, we conduct a Bayesian analysis 
of the event based on the measured event time scale combined with the mass function of 
lens objects and the models of the physical and dynamical distributions of objects 
in the Galaxy.  We use the initial mass function of \citet{Chabrier2003a} for the mass 
function of Galactic bulge objects, while  we use the present day mass function of 
\citet{Chabrier2003b} for disk object.  We note that the adopted mass functions extend 
to substellar objects down to $0.01\ M_\odot$.

\begin{figure}
\includegraphics[width=\columnwidth]{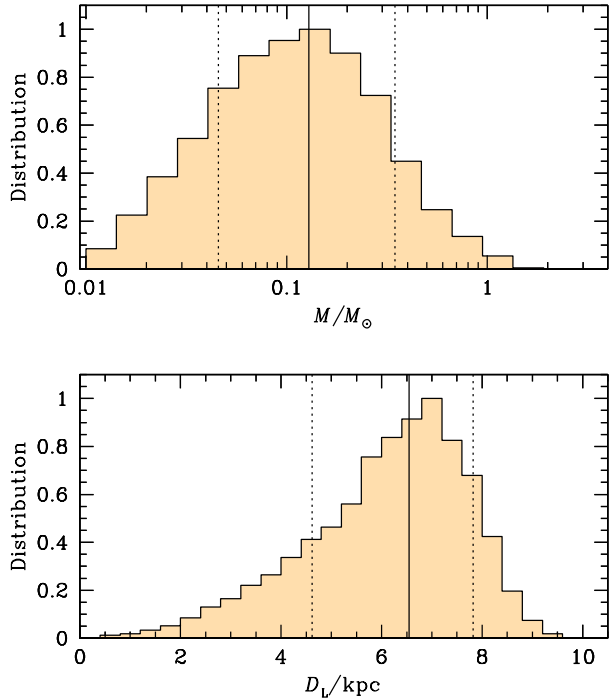}
\caption{
Distributions of the lens mass (upper panel) and the distance to the lens (lower panel) 
estimated by Bayesian analysis.  The solid vertical line in each panel denotes the median 
value and the region surrounded by the dotted lines represents 1$\sigma$ 
(68\%) range of the distribution.  
}
\label{fig:seven}
\end{figure}

\begin{deluxetable}{lc}
\tablecaption{Physical parameters \label{table:four}}
\tablewidth{0pt}
\tablehead{
\multicolumn{1}{c}{Parameter}  &
\multicolumn{1}{c}{Value}  
}
\startdata                                              
 Mass of the primary ($M_1$)         &  $0.13^{+0.21}_{-0.08}$  $M_\odot$   \\
 Mass of the companion ($M_2$)       &  $4.1^{+6.5}_{-2.5}$   $M_{\rm J}$ \\
 Distance to the lens ($D_{\rm L}$)  &  $6.5^{+1.3}_{-1.9}$      kpc        \\
 Projected separation ($a_\perp$)    &  $5.4^{+1.1}_{-1.6}$      au
\enddata                                              
\end{deluxetable}

For the matter density distribution, we adopt the Galactic model of \citet{Han2003}, 
where the matter density distribution is constructed based on a double-exponential disk 
and a triaxial bulge. The velocity distribution is constructed based on the \citet{Han1995} 
model, where the disk velocity distribution is assumed to be Gaussian about the rotation 
velocity of the disk and the bulge velocity distribution is modeled to be a triaxial 
Gaussian with velocity components deduced from the flattening of the bulge via the tensor 
virial theorem. Based on the models, we generate a large number of artificial events by 
conducting a Monte Carlo simulation. We then estimate the ranges of $M$ and $D_{\rm L}$ 
corresponding to the measured event time scale.

In Figure~\ref{fig:seven}, we present the probability distributions of the lens mass 
(upper panel) and distance to the lens (lower panel) obtained from the Bayesian analysis. 
In Table~\ref{table:four}, we also present the estimated masses of the individual lens 
components, $M_1$ and $M_2$, the distance to the lens, $D_{\rm L}$, and the projected 
separation between the lens component, $a_\perp$.  We choose the median values of the 
distributions as representative values and the uncertainties of the physical parameters 
are estimated based on the upper and lower boundaries within which 68\% ($1\sigma$) of 
the distribution is encompassed.

The estimated mass of the primary lens is $M_1=0.13^{+0.21}_{-0.08}\ M_\odot$.  The 
central value corresponds to a low-mass M dwarf, which is the most common lens population.  
The mass of the companion is $M_2=4.1^{+6.5}_{-2.5}\ M_{\rm J}$.  
The upper limit, i.e.\ $\sim 10.6\ M_{\rm J}$, is below 
the deuterium-burning limit of $\sim 13\ M_{\rm J}$, indicating that the companion is 
likely to be a planet. 
The projected separation between the lens 
components is $a_\perp = 5.4^{+1.1}_{-1.6}$ au.  Under the assumption that the snow line, 
which separates regions of rocky planet formation from regions of icy planet formation, 
scales with the mass of a star \citep{Kennedy2008}, the snow line of the host star is 
$a_{\rm sl}=2.7\ {\rm au} (M/M_\odot)\sim 0.35\ {\rm au}$, where 2.7 au is the snow line in 
the Solar System \citep{Abe2000, Rivkin2002}.  If the companion is a planet, then the ratio 
of the $M_1$ -- $M_2$ separation to the snow-line distance of the planetary system is 
$a_\perp/a_{\rm sl}\sim 15.4$.  This ratio corresponds to 
the region beyond Neptune, the outermost planet of the solar system.

\begin{figure}
\includegraphics[width=\columnwidth]{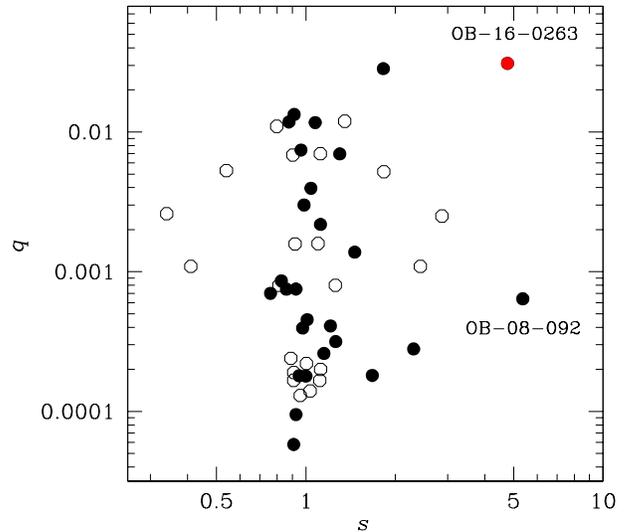}
\caption{
Plot of planet-star separation $s$ vs. the mass ratio $q$ of 48 previously discovered 
microlensing planets. The filled dots represent planets for which the lens parameters 
are uniquely determined, while the empty circles represent planets with close/wide 
degeneracy. For the planets suffering from the degeneracy, we mark two points with 
$s$ and $s^{-1}$. The red-filled dot denotes OGLE-2016-BLG-0263Lb reported in this work.
}
\label{fig:eight}
\end{figure}

\section{Discussion}

The discovery of OGLE-2016-BLG-0263LB demonstrates that high-cadence surveys can provide 
an additional channel of detecting very low-mass companions through repeating events.  
The scientific importance of the repeating-event channel is that the range of planets 
and BDs detectable by microlensing is expanded.

The usefulness of the repeating-event channel is illustrated in Figure~\ref{fig:eight}, 
where we plot the position of OGLE-2016-BLG-0263LB among the 48 previously discovered 
microlensing planets in the $q$-$s$ parameter space. In the plot, filled circles 
represent planets for which the lensing parameters are unambiguous determined. On the 
other hand, empty circles represent planets for which the solutions suffer from degeneracy, 
mostly by the well-known close/wide degeneracy between the solutions with $s$ and $s^{-1}$ 
\citep{Griest1998}. In this case, we mark both solutions.  From the locations of planets, 
it is found that most planets are concentrated in the region around $s=1.0$ \citep{Mroz2017} 
because they were detected from the anomalies that occurred during the lensing magnification 
by their host stars.  By contrast, OGLE-2016-BLG-0263LB is located in the unpopulated region 
of wide separations. It has the largest separation after OGLE-2008-BLG-092LAb, which had a 
projected separation from its host of $s \sim 5.3$ \citep{Poleski2014a}. We note that 
OGLE-2008-BLG-092LAb was also detected through the repeating-event channel.

The repeating-event channel is also important in future space-based microlensing surveys, 
such as {\it WFIRST}, from which  many free-floating planet candidates are expected to be 
detected.  Microlensing events produced by free-floating planets appear as short time-scale 
events.  However, bound planets with large separations from their host stars can also 
produce similar signals, masquerading as free-floating planets \citep{Han2005}.  High-cadence 
ground-based surveys are important because they enable to distinguish some bound planets from 
free-floating planets through the repeating-event channel.  Due to the time-window limit set 
by the orbits of satellites, space-based lensing observations will not observe the bulge field 
continuously.  For example, the {\it WFIRST} survey is planned to be conducted for $\sim 70$ 
days each season.  With the data obtained from space observations, then, it will be difficult 
to sort out short time-scale events produced by bound planets through the repeating-event 
channel.  On the other hand, ground-based surveys continue for much longer periods, $\sim 8$ 
months in average, and thus they can provide an important channel to filter out bound planets 
from the sample of free-floating planet candidates.

\acknowledgments
Work by C.H. was supported by the grant (2017R1A4A1015178) of
National Research Foundation of Korea. 
The OGLE project has received funding from the National Science Centre, Poland, grant 
MAESTRO 2014/14/A/ST9/00121 to A.~Udalski.  OGLE Team thanks Profs.\ M.~Kubiak, G.~Pietrzy{\'n}ski, 
and {\L}.~Wyrzykowski for their contribution to the collection of the OGLE photometric data 
over the past years.
This research has made use of the KMTNet system operated by the Korea
Astronomy and Space Science Institute (KASI) and the data were obtained at
three host sites of CTIO in Chile, SAAO in South Africa, and SSO in
Australia.
Work by A.~Gould was supported by JPL grant 1500811.
A.~Gould and W.~Zhu acknowledges the support from NSF grant AST-1516842. 
Work by J.~C.~Yee was performed under contract with
the California Institute of Technology (Caltech)/Jet Propulsion
Laboratory (JPL) funded by NASA through the Sagan
Fellowship Program executed by the NASA Exoplanet Science
Institute.
We acknowledge the high-speed internet service (KREONET)
provided by Korea Institute of Science and Technology Information (KISTI).
This research has made use of the KMTNet system operated by the Korea
Astronomy and Space Science Institute (KASI) and the data were obtained at
three host sites of CTIO in Chile, SAAO in South Africa, and SSO in
Australia.

\appendix

\begin{deluxetable}{lr}
\tablecaption{Binary-source solution with two timescales \label{table:five}}
\tablewidth{0pt}
\tablehead{
\multicolumn{1}{c}{Parameter}  &
\multicolumn{1}{c}{Value}  
}
\startdata                                              
$t_{0,1}$ (HJD)          &  2457470.465 $\pm$ 0.040  \\ 
$t_{0,2}$ (HJD)          &  2457543.474 $\pm$ 0.040  \\      
$u_{0,1}$                &  0.608 $\pm$ 0.049        \\    
$u_{0,2}$                &  0.394 $\pm$ 0.049        \\ 
$t_{{\rm E},1}$ (days)   &  15.81 $\pm$ 0.82         \\      
$t_{{\rm E},2}$ (days)   &   5.05 $\pm$ 0.53         \\      
$q_F$                    &  0.225 $\pm$ 0.026        \\
$F_s/F_b$                &  3.098/-0.424
\enddata                                              
\end{deluxetable}

In the usual investigation of binary-source solutions
for which the two components are well-separated, these
two components are treated as having fixed separation.   Hence,
in this approximation, the two well-separated events are treated
as having a single Einstein time scale $t_\e$.  Indeed, this is one
of the principal characteristics used to distinguish binary-source
and binary-lens models: if the time scales differ, this implies 
a binary lens with mass ratio $q=(t_{\e,2}/t_{\e,1})^{2}$.

Nevertheless, at some level, the two components must be moving, so that
the Einstein time scales cannot be strictly equal.  Here we quantify
what level of difference is plausible.  Of course it is known that
binary orbital motion can give rise to significant light curve
variations \citep{Han1997} and these can in principle be quite
complicated.   However, here we are working in the wide-separation limit
and so will take a perturbative approach, defined by
\begin{equation}
\epsilon \equiv {\Delta t_\e\over t_\e};
\qquad
\Delta t_\e \equiv t_{\e,2} - t_{\e,1}.
\label{eqa1}
\end{equation}

Since the components are well-separated, $\Delta t_\e$ is sensitive
only to motion along the direction of projected separation 
\begin{equation}
\Delta v_{s,\parallel} = D_s(\mu_{s,2,\parallel} - \mu_{s,1,\parallel})
= D_s\biggl({\theta_\e\over t_{\e,2}} - {\theta_\e\over t_{\e,1}}\biggr)
\simeq -{D_s\theta_\e\Delta t_\e\over t_\e^2}.
\label{eqa2}
\end{equation}
The projected physical separation between the components is
\begin{equation}
a_\perp = D_s\mu_{\rm rel}\Delta t_0 = D_s\theta_\e{\Delta t_0\over t_\e};
\qquad
\Delta t_0 \equiv t_{0,2} - t_{0,1}
\label{eqa3}
\end{equation}
Then, for the system to be bound, $v_{s,\parallel}^2< G M_s/a_\perp$, where
$M_s$ is the total mass of the source (typically $M_s\sim 2\,M_\odot$
for two sources visible in the bulge, although this may not hold
if one of these repeating events is extremely highly magnified).
This can be expressed
\begin{equation}
1> {a_\perp v_{s,\parallel}^2\over GM_s} = 
\biggl({D_s\theta_\e\over t_\e}\biggr)^3{\Delta t_0\over G M_s}\epsilon^2,
\label{eqa4}
\end{equation}
i.e.,
\begin{equation}
{\theta_\e\over {\rm mas}} < 
{t_\e\over D_s({\rm AU}/{\rm kpc})}
\biggl({G M_s\over \epsilon^2\Delta t_0}\biggr)^{1/3}
= {t_\e/{\rm yr}\over D_s/{\rm kpc}}
\biggl({4\pi^2 M_s/M_\odot\over \epsilon^2\Delta t_0/{\rm yr}}\biggr)^{1/3}.
\label{eqa5}
\end{equation}

We now apply this formalism to the case of OGLE-2016-BLG-0263.  We
first search for binary-source solutions as in Section 3.1,
but with the additional degree of freedom $t_\e \rightarrow 
(t_{\e,1},t_{\e,2})$.
The results in Table~\ref{table:five} show that this model comes close to matching
the binary-lens model in terms of $\chi^2$, but at the cost of
a radical divergence of Einstein time scales:
$(t_{\e,1},t_{\e,2})=(15.8,5.0)\,$days.  We note that, in addition, the
blending is negative, $F_b=-0.42$, which corresponds to an $I=19$
``anti-star'', which would require a ``divot'' in the stellar background
of this amplitude.  
Negative blending might be caused either by an incorrect model or 
fluctuation of data for a small $F_b$ case. 
Due to the latter possibility, 
negative blending at this level cannot be excluded.

To apply the formalism, we first note that the flux of the secondary
indicates that it is an upper main sequence star, so that indeed
the masses of the two sources are $M_{s,1} \simeq M_{s,2}\simeq 
1\,M_\odot$.
We then adopt $t_\e = (t_{\e,1} t_{\e,2})^{1/2} = 8.9\,$days, so that
$\epsilon=1.17$, which is outside the ``perturbative regime''.
Nevertheless, if one carries through the non-perturbative calculation,
the final result hardly differs.
We obtain
$$\theta_\e<0.02\,{\rm mas};
\quad \mu<0.8\,{\rm mas\,yr^{-1}}.
$$
The limit on $\mu$ would already make the lens quite unusual, though
hardly unprecedented.  However, the
low value of $\theta_\e$ is more constraining.
For example, for typical bulge lenses
with $D_S-D_L=1\,$kpc, this would imply a lens mass $M_L<0.003\,M_\odot$,
and for disk lenses, $M_L$ would be even lower.  The combination of
somewhat low proper motion and very low Einstein radius
would make this a very remarkable lens.

Moreover, we note that we have been extraordinarily
conservative in putting ``1'' on the r.h.s of Equation~(\ref{eqa4}).  Because
we are viewing only one component of motion and very few systems would
be seen either face-on or near local escape velocity, we could have
chosen a typical value ``1/8'', rather than a strict upper limit.
Thus a more typical source geometry would yield $\theta_\e\sim 0.01\,$mas,
which would imply $M_L<0.0007\,M_\odot$,

We conclude that while the data can be well matched to
a binary-source with large internal motion,  this requires an
improbably small Einstein radius.  Hence, in this case such solutions
are highly disfavored.

%

\end{document}